\documentstyle[prd,aps,preprint,epsfig]{revtex}
\begin{document}

\preprint{ KIAS-P00068}
\date{\today}
\draft

\tightenlines

\title{
Neutrino Interactions In\\
Color-Flavor-Locked Dense Matter\thanks{
Invited talk presented
at the First KIAS Workshop on Astrophysics: Explosive
Phenomena in Compact Astrophysical Systems, KIAS, Seoul, Korea;
24-27 May, 2000.}}

\author{
Deog Ki Hong$^{a,b}$\thanks{E-mail: dkhong@pnu.edu},
Hyun Kyu Lee$^{c}$\thanks{ E-mail: hklee@hepth.hanyang.ac.kr}, Maciej
A. Nowak$^{a,d}$\thanks{E-mail: nowak@kiwi.if.uj.edu.pl} and
Mannque Rho$^{a,e}$\thanks{E-mail: rho@spht.saclay.cea.fr} }

\address{$^a$ School of Physics, Korea Institute for Advanced Study,
Seoul 130-012, Korea\\
$^b$ Department of Physics, Pusan National University,
Pusan, Korea\\
$^c$ Department of Physics, Hanyang University, Seoul, Korea\\
$^d$ Marian Smoluchowski Institute of Physics,
Jagellonian University, 30-059 Krakow, Poland\\
$^e$ Service de Physique Th\'eorique, CE Saclay,
91191 Gif-sur-Yvette, France}

\maketitle

\begin{abstract}
At high density, diquarks could condense in the vacuum with the
QCD color spontaneously broken. Based on the observation that the
symmetry breaking pattern involved in this phenomenon is
essentially the same as that of the Pati-Salam model with broken
electroweak--color $SU(3)$ group, we determine the relevant
electroweak interactions in the color-flavor locked (CFL) phase in
high density QCD. We briefly comment on the possible implications
on the cooling of neutron stars.
\end{abstract}

\section{Introduction}
%\[
%\widehat{a} + \widehat{ab} + \widehat{abc} + \widehat{abcd}
%\]
%
%%\show\frak
%
%\[
%%      {\bf x}^{\bf x} \triangleq z
%      {\bf x}^{\bf x}\triangleq{z} \tensor{T} \frak{E^E}=\frak{mc}^2
%%      {\bf x}^{\bf x}\triangleq {z} \tensor{T} \frak{E}=\frak{mc}^2
%\]
%
%\[
%{\Bbb {NQRZ}} \qquad \because \eth\ggg\bigstar \therefore\blacktriangleright\rightsquigarrow \blacksquare
%\]
%
Recent developments on high-density QCD \cite{highrho}
suggest that diquarks condense in superdense hadronic matter
giving rise to a color-flavor locked state~\cite{wilczek2}.
Among the excitations
on this broken phase vacuum are massive color gluons
metamorphosed to vector mesons and integrally charged quarks
behaving as baryons~\cite{continuity,hrz}. Although such a state
may not be formed in heavy-ion collisions, it may be relevant in
the physics of compact stars such as the cooling of neutron stars.

The ultimate goal of studying the state at high density and at
low temperature is to explore the astrophysical implication of
this broken phase: What is particularly interesting is the role of
neutrinos in the cooling of neutron stars. In preparation for such
a study, we consider how matter in the color-flavor locked state
responds to electroweak (EW) probes. In this talk, we exploit the
observation that the symmetry breaking pattern in the
color-flavor locked (CFL) state is basically the same as that of
the broken color gauge theory, originally proposed by Pati and
Salam~\cite{pati} for a grand unification scheme, to study how
neutrinos (or generally weak current) interact in the CFL matter.

\section{Gauge Theoretical Model For The CFL Phase}
In the minimal Pati-Salam model, $SU(3)_{color}$ gauge group
as well as the flavor group, $SU(2)_L \times U(1)_R$, are
spontaneously broken.  One of the features of this model is the
integrally  charged quarks (i.e., Han-Nambu quark
model~\cite{hannambu}) and the charged massive gluons, both of
which are analogous to the excitations in the color-flavor locked
QCD phase. The gauge bosons get mixed to form mass eigenstates:
for example the neutral gauge bosons, photon $A$, weak boson
$Z_0$, and a combination of gluon $\tilde{V}_0$
%$V_3 + V_8/\sqrt{3}$
are a mixture of the original gauge bosons, $B$ of $U(1)_R$, $W_0$
of $SU(2)_L$ and a combination,  $V_3 + V_8/\sqrt{3}$, of
$SU(3)_{color}$. Consequently it is very natural to expect weak
interactions from the colored objects.  An interesting feature of
this model is that not only photon \cite{hklee} but also neutrinos
(in general, weak current) can probe the color. Despite its
intrinsic elegance, this model has not been considered as a
relevant model for Nature since no experiments performed at low
density have shown any evidence for broken $SU(3)_{color}$ gauge
symmetry.  However recent theoretical developments indicate the
possibility of spontaneous symmetry breaking of $SU(3)_{color}$ in
the hadronic matter at very high density and more intriguingly,
the symmetry breaking pattern -- when suitably reinterpreted -- is
identical to that of the minimal Pati-Salam model. This suggests
that in the color-flavor locked phase, the Pati-Salam model could
be exploited to infer the {\it structure} of electroweak
interactions involving the colored objects. This does not mean
that this is the only way of deriving the interaction forms. As
discussed in \cite{zahedgang}, one can do a systematic
weak-coupling calculations valid at high density taking into
account the color-flavor locking and consequent non-perturbative
effects. The point of this paper is that the symmetry breaking
pattern shares its generic characteristic with other schemes
already available in the literature for different reasons that
allows a simple understanding of the electroweak couplings.

For our purpose we shall choose the gauge group to be $SU(2)_L
\times U(1)_R \times SU(3)_{color}$ as in the minimal version of
Pati-Salam model. One can easily extend it to $SU(3)_L$ or
$SU(4)_L$\cite{pati}.  The basic structure remains the
same.~\footnote{Operationally the symmetry scheme can be extended
to incorporate the left-right global symmetry of nature:
 \begin{eqnarray}
(SU(2)_L\times U(1)_R\times SU(3)_c)_{local}\times (SU(3)_L\times
SU(3)_R)_{global}\\
\rightarrow  (U(1)_{em})_{local}\times (SU(3)_{c+L+R})_{global}.
 \end{eqnarray}}
In addition to the standard Higgs doublet $\phi$, which is
responsible for the electro-weak symmetry breaking transforming
as $(\{2\}, \{1\})$ for $(SU(2)_L, SU(3)_c)$, there are
color-flavored Higgs, $\sigma$, which transform as $(\{2\}+\{2\},
\{3^*\})$
with nonzero vacuum
expectation values for the color flavor diagonal elements:
 \begin{eqnarray}
  \langle \sigma_{aj}\rangle &=& \sigma\ \ {\rm
for} \,\,a=j \ (a=1,2, 3,4,
                \,\, j=1,2,3)\\
             &=& 0 \,\, {\rm otherwise}
\end{eqnarray} where $a=1,2$ and $3,4$ are the doublets of $SU(2)_L$ and $j$
is the color index. In fact,  $a$ are identical to the indices for
global flavor symmetry of the strong interaction when  Cabbibo
mixing is understood. Hence the color-flavored Higgs transforms as
a fundamental representation under the global flavor
transformation. The charm quark with the flavor index $a=4$ is too
heavy for consideration at high density, so we will be primarily
concerned with $SU(3)_f$. The flavor index $a=4$ will not figure
in our discussion. We note that it is analogous to the
color-flavor locked diqaurk condensate :
\begin{equation}
\left<{q_L}_{i\alpha}^a(\vec p){q_L}_{j\beta}^b(-\vec p)\right>=
-\left<{q_R}_{i\alpha}^a(\vec p) {q_R}_{j\beta}^b(-\vec
p)\right>=\kappa(p_F)\epsilon_{\alpha\beta}
\epsilon^{abI}\epsilon_{ijI},
\label{cfl}
\end{equation}
where $\alpha,\beta=1,2$ are Weyl indices, $a,b=1,2,3$ flavor
indices, and $i,j=1,2,3$ color indices. (We will specify the CFL
phenomenon more precisely later.) For now we focus on the
essential point. The diquark  transforms under color as $\{3^*\}$
or $\{6\}$. One can see that $\{3^*\}$~\footnote{It has been
argued that the $\{3^* \}$ scalar channel dominates in the diqaurk
condensate\cite{Alford:1999pa,Pisarski:1999cn,hong}.} is
equivalent to the color-flavored Higgs $\sigma$. This observation
leads us to propose that the symmetry breaking pattern in the CFL
phase can be directly mapped to that in the Pati-Salam model. Of
course we are not implying that the Pati-Salam model is effective
in the zero-density regime.

Let us imagine that both electroweak and color-flavor locking
symmetry breakings have taken place, with $v$ representing the VEV
for the former and $\sigma$ the VEV for the latter. Denote the
electroweak mixing angle (or Weinberg angle) by $\theta_W$ and the
QCD mixing angles by $\beta$ and $\delta$. In the standard
procedure, the gauge covariant couplings of the gauge bosons to
the Higgs scalars give rise to the mass terms for the gauge
bosons. After diagonalizing the mass terms, we have a massless
photon given by~\footnote{For simplicity, we omit the Lorentz
index $\mu$ in the vectors. The charge operator and gauge fields
effective in the CFL phase are denoted by tilde.}
 \begin{equation}
 \tilde{A} = \cos\beta (\sin\theta_W W_3 +
\cos\theta_W B_0) + \sin\beta V_0 \label{a} \end{equation} where \begin{eqnarray}
\tan\beta &=& \frac{2}{\sqrt{3}}\frac{g}{g_s}\sin\theta_W \\
\tan\theta_W &=& \frac{g'}{g}
 \end{eqnarray}
 and
  \begin{eqnarray}
  \tilde{Z} &=& Z  -
\frac{4}{\sqrt{3}}\frac{g_s g}{g^2 + g'^2}
\frac{\sigma^2}{v^2} \cos\theta_W V_0 + {\cal O}(\sigma^2/v^2), \label{z}\\
\tilde{V}_0 &=& -\sin\beta ( \sin\theta_W W_3 + \cos\theta_W B_0) +
\cos\beta V_0 + {\cal O}(\sigma^2/v^2) , \label{v}
\end{eqnarray}
where
\begin{eqnarray}
Z &=&\cos\theta_W W_3 -\sin\theta_W B_0\label{Z0},\\
V_0 &=& \sqrt{\frac{3}{4}}(V_3 + \frac{1}{\sqrt{3}}
V_8).\label{V0} \end{eqnarray}
 Here $g_s$, $g$ and $g^\prime$ are
respectively the gauge coupling constants for $SU(3)_c$,
$SU(2)_L$ and $U(1)$. Now $\tilde{W}^{\pm}$ and $\tilde{V}^{\pm}$
are mixed states of $W^{\pm}$ and $V_{\rho^{\pm}}$: \begin{eqnarray}
\tilde{W}^{\pm} &=& \cos\delta \, W^{\pm} -\sin\delta \, V_{\rho^{\pm}},\\
\tilde{V}^{\pm} &=& \sin \delta \, W^{\pm} + \cos\delta \,
V_{\rho^{\pm}} \end{eqnarray} where
 \begin{equation} \tan \delta =
\frac{g}{g_s}(\frac{M_V}{M_W})^2.
 \end{equation}
The other four colored gauge bosons remain unmixed: \begin{eqnarray}
V^{+}_{K^{*}}, &\,& \, V^{-}_{K^{*}},\\
V^0_{K^{*}}, &\,& \, V^0_{\bar{K}^{*}}\ \ . \end{eqnarray}
 The subscripts
$\rho$, $K^\star$ etc. represent the Higgsed gluons with the
corresponding quantum numbers~\cite{continuity}. The masses of
the gauge bosons are given by
 \begin{eqnarray}
M_V &\sim& g_s\sigma/\sqrt{2},\\
M_W &\sim& gv/\sqrt{2}.
 \end{eqnarray}
Taking $M_V \sim$ few $100$ MeV, we expect that \begin{eqnarray}
\frac{M_V}{M_W} &\sim& 10^{-3}. \end{eqnarray}
 We also expect that
$$\frac{g_s}{g} \sim \frac{g_s}{g'}  \sim 10. $$
The unit of electromagnetic charge is defined as~\footnote{We use
$\tilde{e}$ to distinguish it from the zero-density charge $e$.}
 \begin{equation}
\tilde{e}^2 = \frac{g_s^2g^2\sin\theta_W}{g_s^2 + 4 g^2
\sin^2\theta_W/3}.\label{e}
 \end{equation}
The left-handed quarks and leptons are classified as $(\{2\},
\{3^*\})$  and  $(\{2\}, \{1\})$ respectively.  The right-handed
quarks are classified as $(\{1\}, \{3^*\})$.  Using eqs.(\ref{a})
and (\ref{e}), the electromagnetic charge can be obtained by
 \begin{eqnarray}
\tilde{Q}_{em} &=& Q_{f(lavor)} + Q_{c(olor)}\\
Q_{f} &=& (I_3 + Y/2)_{f}, \label{qem}\\
Q_{c} &=& (I_3 + Y/2)_{c} \end{eqnarray} which gives integer charges of the
Han-Nambu type to the quarks:
\[
\left( \begin{array}{c}
         \tilde{Q}_u \\ \tilde{Q}_d \\ \tilde{Q}_s
       \end{array} \right) = \left( \begin{array}{rcc}
 0 & 1 & 1\\
 -1 & 0 & 0\\
 -1 & 0 & 0

\end{array} \right)
 \label{qemq}
\]
where the (three) columns represent the (three) colors.

In order to identify the charge states of the quarks, which due to
the locking of color with flavor make up a nonet (an octet plus a
singlet), with the baryons in the CFL phase, we have to take the
tensor product of the color and the flavor~\footnote{The simplest
mnemonic for this operation is to arbitrarily assign colors so
that the first column corresponds to a flavor equivalent of
$\bar{u}$, the second column to $\bar{d}$ and the third column to
$\bar{s}$. One would obtain the nonet equivalent to the octet and
singlet of mesons.}. One can identify the two $\tilde{Q}=1$
states with $\tilde{p}$ and $\tilde{\Sigma}^+$, the two
$\tilde{Q}=-1$ states with $\tilde{\Sigma}^-$ and $\tilde{\Xi}^-$
and three $\tilde{Q}=0$ with $\tilde{n}$, $\tilde{\Sigma}^0$ and
$\tilde{\Xi}^0$. The remaining $\tilde{Q}=0$ state is the singlet
baryon which is assumed to be very massive and hence to decouple.

\section{Neutrino Interactions With CFL Excitations}
Given the gauge bosons (\ref{a}), (\ref{z}) and (\ref{v}),
it is straightforward to read off their weak interaction
vertices. We are interested in processes that probe the dense CFL
phase. The electromagnetic interaction may not be useful for
probing the dense phase because the coupling with the matter must
remain strong and most likely lose its memory of the broken phase
when observed by the outside detector. However neutrino
interactions are very weak and can be a good probe.

\subsection{Weak interaction mediated by colored gauge bosons}
Consider the color-gluon annihilation into the lepton pair,
$l \bar{l}$, as an example of weak neutral current interactions:
 \begin{eqnarray}
\tilde{V}^{+} + \tilde{V}^- \rightarrow \tilde{Z} \rightarrow \,
\,\,\, l \bar{l}, \label{zgluon}\\
\tilde{V}^{+} + \tilde{V}^- \rightarrow \tilde{V}_0 \rightarrow \,
\,\,\, l \bar{l}. \label{vgluon}
 \end{eqnarray}
The  coupling at the $\tilde{V}\tilde{V}\tilde{Z}$ vertex in the
process mediated by $\tilde{Z}$, eq.(\ref{zgluon}), is given by
 \begin{eqnarray}
f \cos^2 \delta\, \frac{4}{\sqrt{3}}\frac{g_s g}{g^2 + g'^2}
\frac{\sigma^2}{v^2} \cos\theta_W \sim \frac{4}{\sqrt{3}} g
\cos^3 \theta_W (\frac{M_V}{M_W})^2 \label{sup}
 \end{eqnarray}
which gives a  suppression factor \begin{eqnarray} \sim  (\frac{M_V}{M_W})^2
 \end{eqnarray}
compared to the conventional $\nu\bar{\nu}$ production.
However if there are  substantial amounts of gluon excitations
confined in dense hadronic matter at
nonzero $T$ before it cools down completely, it may overcome the
suppression factor and affect the cooling process
appreciably.

The suppression  factor in the process mediated by $\tilde{V}_0$,
eq.(\ref{vgluon}), due to the vertex $\tilde{V}_0  \,l \bar{l}$ is
given by
 \begin{eqnarray} \sim \sin\beta \,\, \sim \,\, g/g_s. \end{eqnarray}
The propagator in the low energy limit $Q^2 << M_V^2$ is greater
than in eq.(\ref{zgluon}), i.e.,
 \begin{eqnarray}
\frac{1}{Q^2 - M_V^2} \sim \frac{1}{M_V^2}. \end{eqnarray}
 However the amplitude for fusion is
enhanced at the strong interaction vertex,
$\tilde{V}\tilde{V}\tilde{V}_0$, by a factor of $f$, and we get
the factor for the amplitude \begin{eqnarray} \sim \, Q_{f} g
\frac{g}{g_s}\frac{1}{M_V^2} g_s = Q_{f}
\frac{g^2}{M_V^2}\label{fusion} \end{eqnarray}
 with $Q_{f}$ given by
eq.(\ref{qem}). One can now see that the gluon fusion into the
charged flavor $l \bar{l}$ pair is greater than the weak neutral
current by a factor of $\sim (\frac{M_V}{M_W})^{-2} \sim 10^6$ and
hence comparable to photon mediated processes~\cite{hklee}.
However this enhancement does not apply to gluon-mediated $\nu
\bar{\nu}$ processes because $Q_{f}$ is zero for neutrino. In
general, for the neutral current with neutrinos, the contribution
from color-gluon mediated processes in the broken phase vanishes
since the amplitude is proportional to $Q_{f}$(neutrino) which is
$=0$.  We arrive at the same conclusion for $q\bar{q}
\rightarrow  \nu \bar{\nu}$.

The charged current weak interaction
in the process  mediated by $\tilde{V}_0$ is also
comparable to the ordinary weak interaction strength for the
neutrino-quark interaction in the low-energy limit.
Consider the following processes in  matter,
\begin{eqnarray}
q + l &\rightarrow& q' + \nu(\bar{\nu}), \label{qlnu}\\
q &\rightarrow& q' + l + \nu(\bar{\nu}). \label{qnu}
\end{eqnarray}
As in the gluon annihilation processes, there are two amplitudes that
 can be decomposed  into three parts:
 quark gauge boson vertex, propagator, gauge boson-lepton-neutrino vertex,
 \begin{eqnarray}
qq'\tilde{W}^{\pm} \,\,  &\rightarrow& \,  \tilde{W}^{\pm}
\rightarrow  \,\,\,  l\nu\tilde{W}^{\pm}
,  \label{qqz}\\
qq'\tilde{V}^{\pm}\, \,\, &\rightarrow& \,  \tilde{V}^{\pm}\,
\rightarrow \,\,\,  l\nu \tilde{V}^{\pm}. \label{qqv}
 \end{eqnarray}
In the low energy limit, eq.(\ref{qqz}) gives the ordinary  weak
amplitude \begin{eqnarray} \sim \frac{g^2}{M_W^2}.\label{qqza}
 \end{eqnarray}
It is easy to see that the contribution of the color
gauge-boson-mediated process, eq.(\ref{qqv}), also gives an
amplitude comparable to that of the $W^{\pm}$ mediated process,
\begin{eqnarray} \sim g\frac{g}{g_s}(\frac{M_V}{M_W})^2 \frac{1}{M_V^2}
g_s \sim
\frac{g^2}{M_W^2}.\label{qqva}
 \end{eqnarray}

It should be noted however that the quark decay mediated by
$\tilde{V}_0$ in eq.(\ref{qnu}) cannot take place because of the
energy conservation: the quarks with different colors but with
same flavor have the same mass. Therefore the neutrino production
mediated by the color-changing weak current is limited to the
process in eq.(\ref{qlnu})
 \begin{eqnarray} q_r \,\, + \,\, e^- \,\,\,
&\rightarrow& \,\,\, q_b \,\, + \nu
\label{nu}\\
q_b \,\, + \,\, e^+ \,\,\, &\rightarrow& \,\,\, q_r \,\, +
\bar{\nu}. \label{nubar}
 \end{eqnarray}
To keep the system in a color-singlet state in the cooling
process, these processes should occur equally to compensate the
color change in each process. It implies that these processes
depend on the abundance of positrons in the system.  At
finite temperature in the cooling period, it is expected that
there will be a substantial amount of positrons as well as electrons
as long as the
temperature is not far below $\sim MeV$. Of course the additional
enhancement of the neutrino production due to the CFL phase
depends on the abundance of positrons in the system which
depends mainly on the temperature. If confined colored gluons are
present in the matter in the CFL phase, the same amplitude can be
obtained in eq.(\ref{qqv}) when $qq'$ is replaced with $VV'$.

The result obtained above can be summarized as predicting an
enhancement of the effective four-point coupling constant for the
neutrino production process in the low energy limit.  The
enhancement due to the neutrino-color interaction is suppressed by
factors of $e^{-\Delta/T}$ or $e^{-M_V/T}$, since it depends on
the unpaired excitations above gap which can participate into
neutrino-color interaction. Hence for the cooling process at low
temperatures as $\sim 10^9$K it is not so effective. However
during the early stage of  proto-neutron star the temperature is
expected to be high enough  $\sim 20-50$MeV \cite{pons}to see the
effect of the enhancement due to color excitations.

In the next section, the evolution of the effective coupling
constant is used with the help of a renormalization group analysis
to show that there is an additional enhancement of the coupling
constant down to the low temperature.

\subsection{Effective four-point  Fermi coupling constant}
In this subsection, we calculate how the weak coupling
constant runs in dense matter.

In dense matter the gluons are screened since soft gluons can
decay into particles and holes near the Fermi surface. The
one-loop screening effect at zero temperature has been calculated
in the literature~\cite{manuel}. In the high density limit, the
one-loop vacuum polarization density is given, with $M^2=N_f
g_s^2\mu^2/(2\pi^2)$ and $V^{\mu}=(1,\vec v_F),~{\bar V}^{\mu}
=(1,-\vec v_F)$, by
\begin{equation}
\Pi^{\mu\nu}_{ij}(p) =-{iM^2\over2}\delta_{ij} \int{{\rm
d}\Omega_{\vec v_F}\over 4\pi} \left({-2\vec p\cdot\vec
v_FV^{\mu}V^{\nu}\over p\cdot v +i\epsilon \vec p\cdot\vec
v_F}+g^{\mu\nu}-{ V^{\mu}{\bar V}^{\nu}+{\bar
V}^{\mu}V^{\nu}\over2}\right)+{\cal O}(1/\mu).
\end{equation}
The terms involving quark-anti-quark pair creation and gluon loops
are suppressed by $1/\mu$ and are ignored. At low energy,
$p_0\ll|\vec p|\sim p_F$ for the gluon and hence the gluon
propagators take the following form in the Landau
gauge~\cite{igor}:
\begin{equation}
i{\cal D}_{\mu\nu}^{ij}(p_0,\vec{p})\simeq
\delta^{ij} \frac{|\vec{p}|}{|\vec{p}|^3+i\pi M^2 p_0}
O^{(1)}_{\mu\nu}+
\delta^{ij} \frac{1}{p_0^2-|\vec{p}|^2- M^2 }
O^{(2)}_{\mu\nu}.
\label{D-long}
\end{equation}
The projectors for the magnetic and electric modes are
respectively
\begin{eqnarray}
O^{(E)}=P^{\perp}+\frac{(u\cdot p)^2}{(u\cdot p)^2-p^2} P^{u},
&\quad & O^{(M)}=-\frac{(u\cdot p)^2}{(u\cdot p)^2-p^2} P^{u},
\end{eqnarray}
with
\begin{equation}
P^{\perp}_{\mu\nu}= g_{\mu\nu}-\frac{p_{\mu}p_{\nu}}{p^2},
\quad \mbox{and} \quad
P^{u}_{\mu\nu}= \frac{p_{\mu}p_{\nu}}{p^2}
-\frac{p_{\mu}u_{\nu}+u_{\mu}p_{\nu}}{(u\cdot p)}
+\frac{u_{\mu}u_{\nu}}{(u\cdot p)^2}p^2,
\end{equation}
where $u_{\mu}=(1,0,0,0)$. At low energy, the electric gluons are
Debye screened with a screening mass $M$ and the magnetic modes
are dynamically screened (or Landau damped) at
$p_0\ne0$~\cite{son}.

As argued by Alford et al~\cite{wilczek2}, at high density
with three light flavors, the
$SU(3)_c$ gauge symmetry is spontaneously broken by forming a
color-flavor-locked diquark condensate, eq.~(\ref{cfl}).
Then, by Higss mechanism all eight gluons get mass of order of
$g_s\mu$,
which can be easily seen by calculating the one-loop vacuum polarization
tensor with quarks with Majorana mass (or a gap) $\Delta$ generated
by the diquark condensate
~\cite{Son:2000cm,Beane:2000ms,Zarembo:2000pj,Rischke:2000ra}:
\begin{eqnarray}
\Pi_{ij}^{\mu\nu}(p)\!\!\!&=&\!\!\! -g_s^2\int{d^4q\over (2\pi)^4}
{\rm tr}\left[T^A\gamma^{\mu}{q_{\parallel}\cdot\gamma+\Delta\over
q_{\parallel}^2+\Delta^2}T^A\gamma^{\nu}
{(q+p)_{\parallel}\cdot\gamma+\Delta\over (p+q)_{\parallel}^2+\Delta^2}
\right]\\ %+\left(\sigma\leftrightarrow \bar\sigma\right)\\
\!\!\!&\simeq&\!\!\!
0.86{iN_f\over3}{g_s^2\mu^2\over 2\pi^2}\delta_{ij}
\left(g^{\mu\nu}-{p^{\mu}p^{\nu}\over p^2}\right)+\cdots
\end{eqnarray}
where $q_{\parallel}^{\mu}=(q_0,\vec v_F\vec q\cdot\vec v_F)$ and
the ellipses denote terms containing more powers of momentum. At
low momentum all gluons get a dynamical mass,
$M_V\simeq0.2g_s\mu$ for $N_f=3$, independent of the gap, $\Delta$, though
the relevant scale for the dynamical mass generation is of order
of $\Delta$. Let us consider the weak decay of
light quasi-quarks, described by the four-Fermi interaction:
\begin{eqnarray}
{\cal L}_{Fermi}&=&{G_F\over\sqrt{2}} \sum_{\vec v_F}
\bar \psi_L(\vec v_F,x)\gamma^{\mu}\psi_L(\vec v_F,x)
\bar \nu_L(x)\gamma_{\mu}\nu_L(x)\\
&=&{G_F\over\sqrt{2}}\sum_{\vec v_F} \psi_L^{\dagger}(\vec
v_F,x){\psi}_{L} (\vec v_F,x) \bar\nu_L(x)\!
\mathrel{\mathop{v\!\!\!\!/}}\nu_L(x)
\end{eqnarray}
where $G_F=1.166\times10^{-5}~{\rm GeV}^{-2}$ is the Fermi
constant and $\psi$ denotes the quasi-quark near the Fermi surface,
projected from the quark field $\Psi$ as in~\cite{Hong:2000ru},
\begin{equation}
\psi(\vec v_F,x)={1+\vec \alpha\cdot \vec v_F\over2} e^{-i\vec
v_F\cdot \vec x}\Psi(x).
\end{equation}

Since the four-Fermi interaction of quarks with opposite momenta
are marginally relevant and gets substantially enhanced at low
energy, it may have significant corrections to the couplings to
quarks of those weakly interacting particles~\cite{Hong:2000ru}:
\begin{eqnarray}
\delta{\cal L}_{\nu q}&=&{G_F\over\sqrt{2}}
\psi^{\dagger}_{L}(\vec v_F,x)\psi_{L}
(\vec v_F,x) \bar\nu_L(x)\!
\mathrel{\mathop{v\!\!\!\!/}}\nu_L(x)\nonumber\\
& &\times {ig_{\bar3}\over 2 M_V^2}\delta^A_{tv;us}\int_y
\left[\bar \psi_t(\vec v_F^{\prime},y)\gamma^0
\psi_s(\vec v_F^{\prime},y)
\bar\psi_v(-\vec v_F^{\prime},y)\gamma^0\psi_u(-\vec v_F^{\prime},y)
\right]\\
&=&{4\over 3}{g_{\bar 3}\over 2\pi}{G_F\over\sqrt{2}}
\psi^{\dagger}_{L}(\vec v_F,x)\psi_{L} (\vec v_F,x)
\bar\nu_L(x)\bar v\cdot\gamma\nu_L(x), \nonumber
\end{eqnarray}
where $\vec v_F$ and $\vec v_F^{\prime}$ are summed over and $g_{\bar3}$
is the value of the marginal four-quark coupling at the screening mass
scale $M$.
In terms of the renormalization group (RG) equation
at a scale $E$
\begin{equation}
{dG_F(t)\over dt}={4\over 3}{g_{\bar 3}(t)\over 2\pi}G_F(t),
\end{equation}
where $t=\ln E$. The scale dependence of the marginal four-quark
coupling in the color anti-triplet channel is calculated
in~\cite{hong,Hong:2000ru}. At $E\ll\mu$
\begin{equation}
{\bar g}_{\bar3}(t)\simeq {4\pi\over 11}\alpha_s(t).
\end{equation}
Since $\alpha_s(t)=2\pi/(11t)$, we get
\begin{equation}
G_F(E)\simeq G_F(\mu)\left({\mu\over E}\right)^{{16\pi\over 363}}.
\end{equation}
Since the RG evolution stops at scales lower than the gap,
the low energy effective Fermi coupling in dense matter is therefore
\begin{equation}
G_F^{\rm eff}=G_F\cdot\left({\mu\over \Delta}\right)^{{16\pi\over 363}}.
\end{equation}
We emphasize that this enhancement applies equally to the $\beta$
decay of quarks and other neutrino production processes described in the
previous section.

\section{Neutron-Star Cooling Processes}
At asymptotic density and low temperature ($T\ne0$), the
relevant excitations are quasi-quarks that are not Cooper-paired,
and 17 Nambu-Goldstone bosons. All other massive particles,
Higgsed gluons and other massive excitations~\cite{RSWZ} are
expected to be out of thermal equilibrium and
decoupled.\footnote{If density is not too high, that is, in the
regime relevant for such compact stars as neutron stars, there
may take place Goldstone boson condensation from the top-down
point of view as from the bottom-up. There may also be
excitations of generalized (bound-state) mesons discussed in
\cite{RSWZ} that become low-lying and hence participate in the
cooling process. We cannot address these issues in this paper.}
Thus the main cooling processes must be the emission of weakly
interacting light particles like neutrinos or other
(weakly interacting) exotic light
particles ({\it e.g.} axions) from the quasi-quarks and
Nambu-Goldstone bosons in the thermal bath.

We can think of two processes. The first one is the weak decay of
quasi-quarks considered in the previous subsection where the
running weak coupling indicates a modest enhancement of the
process. Since neutrinos interact weakly, it can effectively
carry away the energy of quark matter. For the neutrino emissivity
from quasi quarks, the so-called Urca process is relevant.
The neutrino emissivity by
the direct Urca process in quark matter,
which is possible for most cases
in quark matter, was calculated by Iwamoto~\cite{Iwamoto:1980eb}.
For the CFL superconcductor,
we expect the calculation goes in
parallel and the neutrino emissivity is
\begin{equation}
\epsilon_{\rm direct}\propto\alpha_s\rho Y_e^{1/3}T^6,
\end{equation}
where $\rho$ is the density, $T$ is the temperature of
the quark matter, and
$Y_e$ is the ratio between the electron and baryon density.
On the other hand, the neutrino
emissivity by the modified Urca process, which is the dominant
process in the standard cooling of neutron stars~\cite{friman},
is suppressed
by $\left(\Delta/\mu\right)^4$, since the pion coupling to quarks
is given by $g_{qq\pi}\sim \Delta/\mu$~\cite{Hong:2000ei}.
Thus, the neutrino emmisivity by the modified Urca process in the
CFL quark matter is greatly suppressed in the CFL quark matter,
compared to normal quark matter.
Futhermore, since the pion-pion interaction in the CFL quark
matter are also suppressed by
$\Delta/\mu$~\cite{Son:2000cm,Rho:2000xf}, we note that all the
low energy excitations in the CFL quark matter are extremely
weakly coupled. But, since most excitations in the CFL quark matter
are gapped and frozen out, the CFL quark matter has a quite small
heat capacity and cools down very slowly at temperatures
lower than the gap~\cite{Alford:2000sx}.

The second process that appears to be important is the
bremsstrahlung emission of neutrino pairs from massless colored
Nambu-Goldstone bosons: $\phi\phi\to \phi\phi\nu\bar\nu$.

The relevant terms {\it before} the gauge boson mass matrix is
diagonalized can be written as
 \begin{eqnarray}
{\cal L}^{int} = \,\, \cdots + \,\, \frac{g_s}{2}
{\phi^{\dagger}}^i_{a^{\prime}}
\stackrel{\leftrightarrow}{\partial_{\mu}}\phi_{a^{\prime}}^j
(\lambda^A)_{ij} V_{\mu}^A  \,\, + \,\, {g\over
\cos\theta_W}Z_{\nu}\left(T_3-\sin^2\theta_WQ_f\right)_{ab} {\bar
l}_{La}\gamma^{\nu}l_{Lb} \,\, + \cdots \label{phinu}
\end{eqnarray}
where $\phi$'s are the scalar Nambu-Goldstone bosons and
$l_L=(\nu,e)_L^T$ is the left-handed lepton isospin doublet. Then
the coupling for  $\phi\phi\nu\bar\nu$ is induced by the
$\tilde{Z}$ exchange
 \begin{eqnarray}
(\phi \,  \phi \, \tilde{Z}) \,\, G_{\tilde{Z}} \, \,
(\nu\bar{\nu}\tilde{Z})
 \end{eqnarray}
where $G_{\tilde{Z}}$ is the $\tilde{Z}$ propagator.  This can be
written as an effective vertex given by \begin{eqnarray} {\cal L}^{eff} =
\cdots + -\frac{4}{\sqrt{3}} {g_sg\over g^2+{g^{\prime}}^2} \cos
\theta_W {\sigma^2\over v^2} \frac{g_s}{2} {1\over M_W^2}{g\over
\cos\theta_W} {\phi^{\dagger}}^i_{a^{\prime}}
\stackrel{\leftrightarrow}{\partial_{\mu}}
(\lambda^8)_{ij}\phi_{a^{\prime}}^j{\bar\nu}_L\gamma^{\mu}\nu_L
  \,\, + \cdots \label{leff}
 \end{eqnarray}
A similar result was given in eq.(\ref{sup}) for
$V^+V^-\nu\bar\nu$. The result (\ref{leff}) can be easily
understood by noting that the Golstone bosons in eq.(\ref{leff})
are nothing but the longitudinal components of the massive
gluons. This process is again suppressed by a factor of $\sim
(M_V/M_W)^2$ compared to the  conventional $\nu\bar\nu$
production.

\section{Discussion and Summary}
In this talk, we argue that the symmetry pattern of the
color-flavor-locked phase of QCD at high density in the presence
of electroweak interactions is mapped to the Pati-Salam model of
grand unification. Then, we have shown that this is a simple way of
deducing the electroweak coupling of the CFL degrees of freedom.
We find that the neutrino interaction with matter in the
color-flavor locked phase can be enhanced by additional
gluon-mediated processes. It remains to be verified that one can
arrive at the same result in the weak-coupling QCD calculation of
the type performed in \cite{zahedgang,RSWZ}.

It is perhaps useful to further comment on the idea of mapping the
EW responses of the CFL phase to the Pati-Salam model. The
symmetry (breaking) pattern is presumably encoded in the effective
potential of the scalars $\phi$ and $\sigma$, ${\cal
V}^{eff}(\phi, \sigma)$.  At low density, the physical vacuum
has minimum at $\langle\phi\rangle \neq 0$,
$\langle\sigma\rangle =0$ and defines
the electro-weak gauge theory, i.e., the established Standard
Model. At some high density, however, the effective potential
could develop a VEV of $\sigma$, as suggested by model
calculations~\cite{highrho}. One of the possibilities is the
color-flavor-locked phase considered here. (Other possibilities
discussed in the literature can also be addressed similarly.)
There are two issues to be resolved in the symmetry breaking
schemes with scalar particles. As in the electro-weak theory at
zero density, there are massive Higgs particles for which the
effective potential is yet to be derived or explained. Presumably
the effect of density might be marginal for these excitations. As
for the color-flavor symmetry breaking, while there is rather
compelling renormalization-group-flow argument to suggest that the
vacuum expectation value of $\sigma$ is non-zero at some
(asymptotic) density, the explicit form of the effective potential
(with density dependence) is yet to be derived.

The essential feature of the neutrino production (except for the
fusion processes) is that the color changes as neutrinos(or
anti-neutrinos) are produced by the processes mediated by the
color gluon exchange.  One might therefore think that such
processes are forbidden due to the color singlet requirement of
the system. However at finite temperature the presence of positron
excitations in the system induces processes which preserve the
color singlet status as explained in the text. It is noted that
the neutrino-color interaction is suppressed at low temperature
cooling stage of neutron star but is expected to be effective at
the early stage of proto neutron star.

 Together with the general
enhancement of the effective four-point coupling constant in RG
analysis,  the enhancement of the neutrino production implies that
the cooling process speeds up as the CFL phase sets in dense
hadronic matter near the critical temperature. But, at temperature
much below the critical temperature, the interaction of quasi-quarks
and pions and kaons is extremely weak, suppressed by
$\Delta/\mu$, and the CFL quark matter cools down extremely slowly.

For a realistic calculation of the cooling rate of compact stars,
we need to also consider the neutrino propagation in the CFL
matter before the neutrinos come out of the system. A recent
study~\cite{cr} suggests that the presence of the CFL phase can
accelerate the cooling process because neutrino interactions with
matter are reduced in the presence of a superconducting gap
$\Delta$. However this result is subject to modification by the
effect of additional interactions -- not taken into account in
this work -- mediated by the colored gluons on the quark
polarization. It would be interesting to see how the enhancement
of the neutrino production correlates with the neutrino-medium
interaction. This is one of the physically relevant questions on
how the enhanced neutrino interaction could affect
neutron-star(proto neutron star) cooling following supernova
explosion. This issue is currently under investigation.

\begin{acknowledgments}
%\subsection*{Acknowledgments}
We wish to thank Mark Alford for useful comments.
We would like to acknowledge the hospitality of KIAS where
this work was initiated as a part of KIAS program for
astro-hadron physics. The work by D.K.H. is supported by
KOSEF grant number 1999-2-111-005-5 and HKL is supported in part
by the interdisciplinary research program of the KOSEF,
grant no. 1999-2-003-5.
\end{acknowledgments}

\end{document}